\documentclass[12pt]{article}
\usepackage{rotate}
\usepackage{epsfig}
\usepackage{psfrag}
\usepackage{a4}

\begin{document}

\newcommand \be  {\begin{equation}}
\newcommand \bea {\begin{eqnarray} \nonumber }
\newcommand \ee  {\end{equation}}
\newcommand \eea {\end{eqnarray}}

\title{\bf Fluctuations and response in financial markets: the subtle nature of
`random' price changes}

\author{Jean-Philippe Bouchaud$^{\dagger,*}$, Yuval Gefen$^{\times}$,\\
 Marc Potters$^*$,
Matthieu Wyart$^{\dagger}$}
\maketitle
{\small
{$^\dagger$ Commissariat \`a l'Energie Atomique, Orme des Merisiers}\\
{91191 Gif-sur-Yvette {\sc cedex}, France}\\

{$^*$ Science \& Finance, Capital Fund Management, 109-111 rue Victor Hugo}\\
{92 532 Levallois {\sc cedex}, France}\\

{$^\times$ Condensed Matter Physics Department, Weizmann Institute of Science,}\\
{ 76 100
Rehovot, Israel}\\
\date{\today}
}

\begin{abstract}
Using Trades and Quotes data from the Paris stock market, we show 
that the random walk nature of traded prices results from a very
delicate interplay between two opposite tendencies: long-range correlated 
market orders that lead to super-diffusion (or persistence), and 
mean reverting limit orders that lead to sub-diffusion (or anti-persistence).
We define and study a model where the price, at any instant, is the result of the 
impact of all past trades, mediated by a non constant `propagator' 
in time that describes the 
response of the market to a single trade. Within this model, the market is 
shown to be, in a precise
sense, at a critical point, where the price is purely diffusive and the average response 
function almost constant. We find empirically, and discuss theoretically, 
a fluctuation-response 
relation. We also discuss the fraction of truly informed market orders, that 
correctly anticipate short term moves, and find that it is quite small.
\end{abstract}


\pagebreak

\section{Introduction}

The Efficient Market Hypothesis (EMH) posits that all available information is 
included in prices, 
which emerge at all times from the consensus between fully rational agents, that would
otherwise immediately arbitrage away any deviation from the fair price \cite{Fama,Samuelson}. 
Price changes can then
only be the result of un-anticipated news and are by definition totally unpredictable.
The price is at any instant of time the best predictor of future prices. One of the central
predictions of EMH is thus that prices should be random walks in time which  (to a good 
approximation) they indeed are. This was interpreted early on as a success of EMH. However, as 
pointed out by Schiller, the observed volatility of markets is far too high to be compatible 
with the idea of fully rational pricing \cite{Schiller}. The frantic activity 
observed in financial markets is another problem: on liquid stocks, there is typically one
trade every 5 seconds, whereas the time lag between of relevant news is certainly much larger. 
More fundamentally, the assumption of rational, perfectly informed agents seems intuitively 
much too strong, and has been criticized by many \cite{Arthur,Shleifer,Orlean}. Even the very concept 
of the fair price of a company appears to be somewhat dubious. 

There is a model at the other extreme of the spectrum where prices also follow a pure random walk,
but for a totally different reason. Assume that agents, instead of being fully rational,
have zero intelligence and take random decisions to buy or to sell, but that
their action
is interpreted by all the others agents as potentially containing some information. Then, 
the mere fact of buying (or selling) typically leads to a change of the ask $a(t)$ (or bid $b(t)$) 
price and hence of a change of the midpoint $m(t)=[a(t)+b(t)]/2$. In the 
absence of reliable information about the `true' price, the new midpoint is 
immediately adopted by all other market participants as the new reference price around which 
new orders are launched. In this case, the 
midpoint will also follow a random walk (at least for sufficiently large 
times), even if trades are not motivated by any rational 
decision and devoid of meaningful information.\footnote{That 
this simplistic model also leads to a random walk behaviour 
for prices has also 
very recently been pointed out in \cite{Hopman}.} This alternative, random trading model has
been recently the object of intense scrutiny, in particular as a simplified approach to the 
statistics of order books \cite{Bak,Maslov,Challet1,Slanina,Farmer1,BMP,Farmer2,Challet2}. Since the order flow is a Poisson process, this assumption 
is quite convenient and leads to tractable analytical models \cite{Farmer2,ustocome}. Perhaps surprisingly, many 
qualitative (and sometimes quantitative) properties of order books can be predicted using 
such an extreme postulate \cite{Farmer1,BMP,Farmer2,Farmer3}. 

Of course, reality should lie somewhere in the middle: clearly, the price cannot wander 
arbitrarily far from a reasonable value, and trades cannot all be random. The interesting 
question is to know which of the two pictures is closest to reality and can be taken as 
a faithful starting point around which improvements can be perturbatively added. 

In this paper, we want to argue, based on a series of detailed empirical results obtained 
on trade by trade data, that the random walk nature of prices is in fact highly non trivial and 
results from a fine-tuned competition between two populations of traders, liquidity 
providers (`market-makers') on the one hand, and liquidity takers (sometimes called 
`informed traders', but see the discussion in Section 4). For reasons that we explain in more
details below, liquidity providers act such as to create anti-persistence (or mean reversion)
in price changes that would 
lead to a sub-diffusive behaviour of the price, whereas liquidity 
takers' action leads to long range persistence and super-diffusive behaviour. Both 
effects very precisely compensate and lead to an overall diffusive behaviour,
at least to a first approximation, such that (statistical) arbitrage opportunities are absent, as expected. However, one can spot out the vestiges of this subtle 
compensation 
from the temporal structure of the market impact function 
(which measures how a given trade affects on average future prices).

The organization of this paper is as follows. We first present (Section 2) our empirical 
results on the statistics of trades, market impact and fluctuations. 
We show in particular that the order flow exhibits long range 
autocorrelations in time, but that this does not lead to any predictability 
in price changes, as also recently noticed in \cite{Hopman}. 
Then, we introduce in 
Section 3 a simple model 
that expresses the price as a linear superposition of the impact of each trade. We 
show that this model allows to rationalize our empirical findings, provided a specific
relation between the temporal autocorrelation of the sign of the trades (i.e. buyer 
initiated or seller initiated) and the temporal response to a single trade is satisfied. 
Finally, in Section 4, we give intuitive arguments that 
allow one to understand the market forces at the origin of this subtle balance between 
two opposite effects, 
which dynamically leads to absence of statistical arbitrage opportunities. 
We argue that
in a very precise sense, the market is sitting on a critical point; 
the dynamical compensation of two conflicting tendencies is similar to
other complex systems such as the heart \cite{heart}, driven by two 
antagonist systems (sympathetic and para-sympathetic), or certain 
human tasks, such as balancing of a long stick \cite{balancing}. 
The latter example illustrates very clearly the idea of dynamical 
equilibrium, and shows how any small deviation from perfect balance 
may lead to strong instabilities. This near instability may well be at the 
origin of the fat tails and volatility clustering observed in financial data 
(see e.g. \cite{Lux,Stanley,Stanley-Gopi,Book,Cont,Muzy}). Note 
that these two features are indeed present in the `balancing stick' 
time series studied in \cite{balancing}.   

\section{Market impact and fluctuations}

\subsection{Presentation of the data and definitions}

In this study, we have analyzed trades and quotes data from liquid French stocks
in the years 2001 and 2002, although qualitatively similar results were also obtained 
on British stocks as well. The advantage of the French market, however, is that it is 
fully electronic whereas only part of the volume is traded electronically in the London
stock exchange. We will illustrate our results mainly using the France-Telecom stock,
which is one of the most actively traded stocks, for which statistics are particularly
good.   

There are two data files for each stock: one gives the list of all successive {\it quotes},
i.e. the best buy (bid, $b$) and sell (ask, $a$) prices, together with the available volume, and the
time stamp accurate to the second. A quote can change either as a result of a trade, or 
because new limit orders appear, or else because some limit orders are canceled. The 
other data file is the list of all successive {\it trades}, with the traded price, traded 
volume and time stamp, again accurate to the second. Sometimes, several trades are recorded
at the very same instant but at different prices: this corresponds to a market order of a 
size which exceeds the available volume at the bid (or at the ask), and hits limit orders deeper 
in the order book. In the following, we have grouped all these trades together as a single 
trade. This allows one to create chronological sequences of trades and quotes, such that
between any two trades there is at least one quote.  

The last quote before a given trade allows one to define the sign of each trade: if
the traded price is above the last midpoint $m=(a+b)/2$, this means that the 
trade was triggered by a market order (or marketable limit order) to buy, and we will
assign to that trade a variable $\varepsilon=+1$. If, one the other hand the traded price is
below the last midpoint $m=(a+b)/2$, then $\varepsilon=-1$.
With each trade is also associated a volume $V$, corresponding to the total number of shares
exchanged. 

Trades appear at random times, the statistics of which being itself non trivial (there 
are intra-day seasonalities and also clustering of the trades in time). We will not be
interested in this aspect of the problem and always reason in terms of trade time, i.e.
time advances by one unit every time a new trade (or a series of simultaneous trades) is
recorded. We have also systematically discarded the first ten and the last ten minutes of
trading in a given day, to remove any artifacts due to the opening and closing of the market.
Many quantities of interest in the following are two-time observables, that is, compare 
two observables at (trade) time $n$ and $n + \ell$. In order to avoid overnight effects, 
we have restricted our analysis to intra-day data, i.e. both $n$ and $n + \ell$ belong to the
same trading day. We have also assumed that our observables only depend on the time lag $\ell$.

On the example of France-Telecom, on which we will focus mostly, there are on the order of 
10 000 trades per day. For example, the total number of trades on France-Telecom during 2002 
was close to  $2. \, 10^6$; this allows quite accurate statistical estimates of various quantities. The volume 
of each trade was found to be roughly log-normally distributed, with $\langle \ln V \rangle 
\simeq 5.5$ and a root mean square of $\Delta \ln V \simeq 1.8$. 
The range of observed values of 
$\ln V$ is between $1$ and $11$. 

\subsection{Price fluctuation and diffusion}

The simplest quantity to study is the average mean square fluctuation of the price between 
(trade) time $n$ and $n + \ell$. Here, the price $p_n$ is defined as the mid-point before the 
$n$th trade: $p_n \equiv m_{n^-}$. In this paper, we always consider detrended prices, 
such that the empirical drift is zero. We thus define ${\cal D}(\ell)$ as:
\be
{\cal D}(\ell) = \left \langle \left(p_{n+\ell}-p_n\right)^2 \right \rangle.
\ee
As is well known, in the absence of any linear correlations between successive price changes,
${\cal D}(\ell)$ has a strictly diffusive behaviour, i.e.
\be
{\cal D}(\ell) = D \ell,
\ee
where $D$ is a constant. In the presence of short-ranged correlations, one expects 
deviations from this behaviour at short times. However, on liquid
stocks with relatively small tick sizes such as France-Telecom (FT), 
one finds a remarkably 
linear behaviour for ${\cal D}(\ell)$, even for small $\ell$. 
The absence 
of linear correlations in price changes is equivalent to saying that (statistical) arbitrage
opportunies are absent, even for high frequency trading. 
In fact, in order to emphasize 
the differences from a strictly diffusive behaviour, we have studied 
the quantity 
$\sqrt{{\cal D}(\ell)/\ell}$ (which has the dimension of Euros). 
We show this quantity 
in Fig. 1 for FT, averaged over three different periods: first semester of 2001 (where the 
tick size was $0.05$ Euros), second semester of 2001, and the whole of 2002 (where the tick 
size was $0.01$ Euros). One sees that ${\cal D}(\ell)/\ell$ is indeed nearly constant, with a 
small `oscillation' on which we will comment later. Similar plots can be observed for other
stocks (see Fig. 2). We have noted that for stocks with larger ticks, a slow decrease of 
${\cal D}(\ell)/\ell$ is observed, corresponding to a slight anti-persistence (or sub-diffusion) 
effect.  

\begin{figure}
\begin{center}
\psfig{file=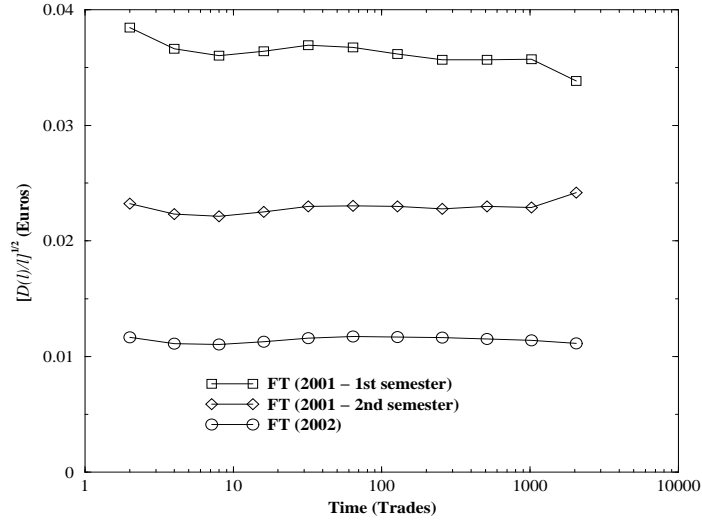,width=7cm,height=9.2cm,angle=270} 
\end{center}
\caption{Plot of $\sqrt{{\cal D}(\ell)/\ell}$ as a function of $\ell$ for France-Telecom, 
during three different periods. The variation of ${{\cal D}(\ell)/\ell}$ 
with $\ell$ is very small, in particular in the small tick ($0.01$ Euros) period 
(July 2001 -- December 2002). For the large tick size period ($0.05$ Euros; January 2001
-- June 2001), there is a systematic downward trend: see also Fig. 2.}
\label{Fig1}
\end{figure}

\begin{figure}
\begin{center}
\psfig{file=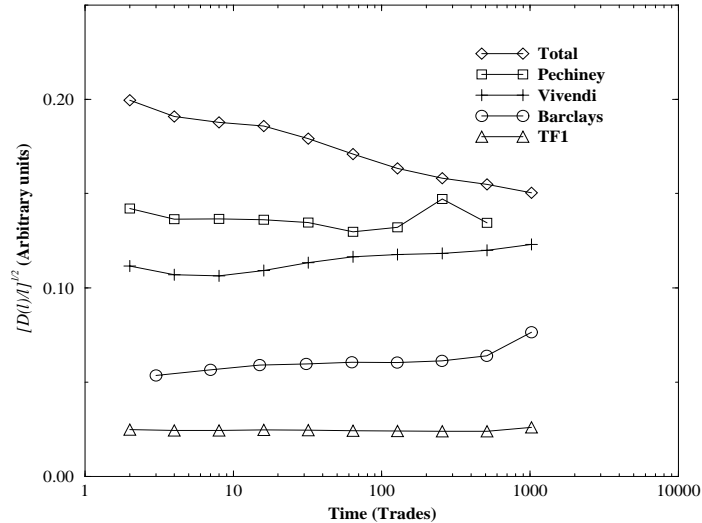,width=7cm,height=9.2cm,angle=270} 
\end{center}
\caption{Plot of $\sqrt{{\cal D}(\ell)/\ell}$ as a function of $\ell$ for other 
stocks
during the year 2002, except Barclays (May-June 2002).
The $y$-axis has been rescaled arbitrarily for clarity. 
We note that stocks with larger tick size 
tend to reveal a stronger mean-reverting effect.}
\label{Fig2}
\end{figure}

The conclusion is that the random walk (diffusive) behaviour of stock prices appears even at
the trade by trade level, with a diffusion constant $D$ which is of the order 
of the typical bid-ask squared. From Fig. 1, one indeed sees that $\sqrt{{\cal D}(1)}
\sim 0.01$ Euros, which is precisely the tick size, and FT has a typical bid-ask 
spread equal to one or two ticks. This coincidence is interesting. It might 
suggests that price changes are to a large extent induced by the trading 
activity itself, independently 
of real news (unless of course if the news flow is 
itself on the scale of seconds 
and that each news item has an impact on the price that is commensurate to the 
bid-ask spread). Much stronger arguments in favor of this point,
based on estimates of the fraction of informed trades, will be given 
below. This conclusion seems to 
imply that the price may, on the long run, wander arbitrarily far from the 
fundamental price, which would be absurd. However, even if one assumes that 
the fundamental price is independent of time, a typical $3 \%$ noise induced 
daily volatility would lead to a significant (say a factor 2) 
difference between the traded price and the fundamental price only after a 
few years \cite{wyart}. Since the fundamental price of a company is probably 
difficult to determine better than within a factor two, say (see e.g. 
\cite{Shleifer,Black}), 
one only expects fundamental effects to be relevant on very long time scales (as 
indeed suggested by the empirical results of de Bondt and Thaler \cite{Thaler}), but that
these are totally negligible on short (intra-day) time scales of interest here. We will in fact
see below (cf. Eq. (\ref{meanrevert}) 
that the reference price that market participants seem to have in mind is 
in fact a short time average of the past price itself, rather than 
any fundamental price.

\subsection{Response function and market impact}

In order to better understand the impact of trading on price changes, one can
study the following {\it response function} ${\cal R}(\ell)$, defined as:
\be
{\cal R}(\ell) = \left \langle \left(p_{n+\ell}-p_n\right) \cdot \varepsilon_n \right \rangle,
\ee
where $\varepsilon_n$ is the sign of the $n$-th trade, introduced in Section 2.1. The 
quantity ${\cal R}(\ell)$ measures how much, on average, the price moves up 
conditioned to a buy order at time $0$ (or
a sell order moves the price down) a time $\ell$ later. As will be clear below,this quantity is  however not the market response to a single trade, a
quantity that will later be denoted by $G_0$. 
A more detailed object can in fact
be defined by conditioning the average to a certain volume $V$ of the $n$-th trade:
\be
{\cal R}(\ell,V) = \left. 
\left \langle \left(p_{n+\ell}-p_n\right) \cdot \varepsilon_n \right \rangle \right|_{V_n = V}.
\ee
Previous empirical studies have mostly focused on the volume dependence
of ${\cal R}(\ell,V)$, and established that this function is strongly 
concave as a function of the volume \cite{Hasbrouck,Barra,Korn,Gopi,Hopman}. 
In \cite{Lillo}, a thorough analysis of U.S. stocks was analyzed in terms of a piecewise  
power-law dependence for ${\cal R}(\ell=1,V) \propto V^\alpha$, with
an exponent $\alpha \simeq 0.4$ for small volumes, and a smaller value 
($\alpha \simeq 0.2$) for larger volumes. 
In a previous publication \cite{Bali}, some of us 
have proposed that this dependence might in fact be logarithmic
(see also a footnote in \cite{Gopi}): ${\cal R}(\ell=1,V) = R_1 \ln V$ (where $R_1$ is a stock dependent constant), a 
law that seems to satisfactorily account for
all the data that we have analyzed. The empirical determination of the temporal structure 
of ${\cal R}(\ell,V)$ has been much less investigated (although one can find
in \cite{Gopi} somewhat related results on a coarse-grained version of  ${\cal R}(\ell,V)$). Preliminary empirical 
results, published in \cite{Bali}, reported that ${\cal R}(\ell,V)$ 
could be written in a factorized form (first suggested on theoretical 
grounds in \cite{Farmer1}): 
\be\label{factor}
{\cal R}(\ell,V) \approx {\cal R}(\ell) f(V); \qquad f(V) \propto \ln V,
\ee
where ${\cal R}(\ell)$ is a slowly varying function 
that initially increases up to $\ell \sim 100-1000$ and then is seen to decrease back, with 
a rather small overall range of variation. The initial increase of $R(\ell)$ 
was reported in \cite{Hasbrouck} and has also 
recently been noticed by Lillo and Farmer \cite{Farmer3}. Here, we provide much better data 
that supports
both the above assertions. We show for example in Fig. 3 the temporal structure of 
${\cal R}(\ell)$ for France Telecom, for different periods. Note that ${\cal R}(\ell)$ 
increases by a factor $\sim 2$ between $\ell=1$ and $\ell=\ell^* \approx 1000$, before decreasing 
back. Similar results have been obtained for many different stocks as well: 
Fig. 4 shows a small selection of other stocks, where the non monotonous behaviour 
of ${\cal R}(\ell)$ is shown. However, in some cases (such as Pechiney), the maximum
is not observed. One possible reason is that the number of daily trades is in this case 
much smaller ($\sim 1000$), and that $\ell^*$ is beyond the maximum intra-day time 
lag.    

\begin{figure}
\begin{center}
\psfig{file=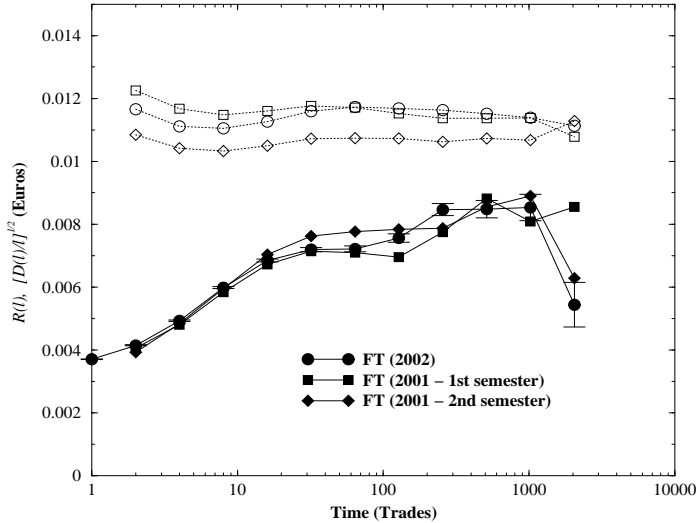,width=7cm,height=9.2cm,angle=270}
\end{center}
\caption{Average response function ${\cal R}(\ell)$ for FT, during three different periods (black
symbols). We have given error bars for the 2002 data. For the 2001 data, 
the $y-$axis has been rescaled to best collapse onto the 2002 data. 
Using the same rescaling factor, we have also shown the data of Fig. 1. The fact that the same
rescaling works approximately for ${\cal D}(\ell)$ as well 
will be dwelled further in Section 2.4 below.}
\label{Fig3}
\end{figure}

\begin{figure}
\begin{center}
\psfig{file=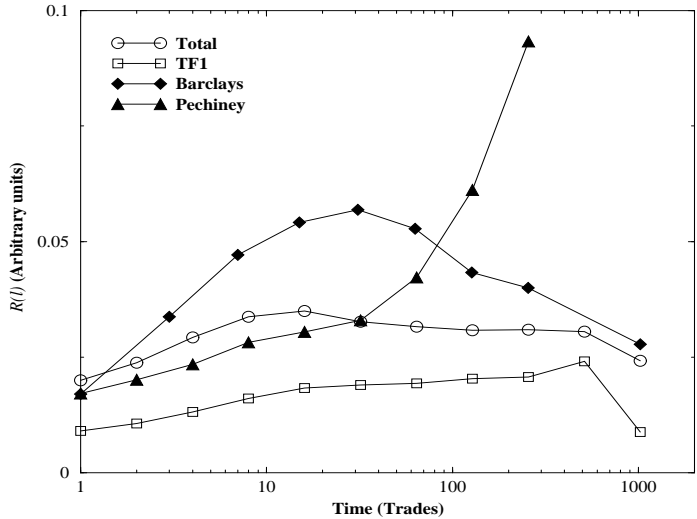,width=7cm,height=9.2cm,angle=270} 
\end{center}
\caption{Average response function ${\cal R}(\ell)$ for a restricted 
selection of stocks, during the year 2002.}
\label{Fig4}
\end{figure}

The existence of a time scale $\ell^*$ beyond which 
${\cal R}(\ell)$ decreases is thus both statistically significant, 
and to a large degree independent of the considered stock. 
On the other hand, the amplitude 
of the change of ${\cal R}(\ell)$ seems to be stock dependent. 
As will be clear later, 
the slowly varying nature of ${\cal R}(\ell)$ and the fact that 
this quantity reaches a maximum are non trivial results that 
will require a specific interpretation. 

Turning now to the factorization property of ${\cal R}(\ell,V)$, Eq. (\ref{factor}), we
illustrate its validity in Fig. 5, where ${\cal R}(\ell,V)/f(V)$ is plotted as a 
function of $\ell$ for different values of $V$. The function $f(V)$ was chosen for 
best visual rescaling, and is found to be close to $f(V)=\ln V$, as expected. Note that 
for the smallest volume (open circles), the long time behaviour of ${\cal R}(\ell,V)$ 
seems to be different, which is probably due to the fact that small volumes 
are in fact more likely to be large volumes chopped up into small pieces. 

\begin{figure}
\begin{center}
\psfig{file=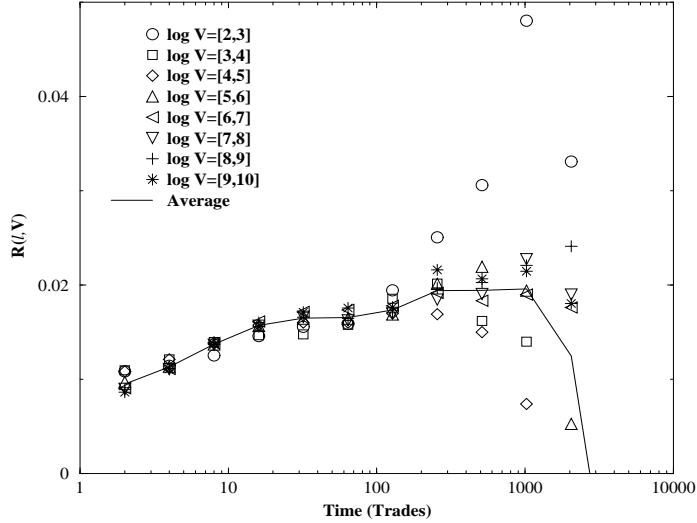,width=7cm,height=9.2cm,angle=270} 
\end{center}
\caption{Average response function ${\cal R}(\ell,V)$, conditioned to a certain volume $V$, 
as a function of $\ell$. Data for different $V$'s have been divided by 
$f(V) \propto  \ln V$ 
such as to obtain good data collapse. The thick line corresponds to ${\cal R}(\ell)$ (unscaled).}
\label{Fig5}
\end{figure}

One has to keep in mind that the response function ${\cal R}(\ell)$ 
captures a small
systematic effect that relates the average price change to the sign of 
a trade.
However, the fluctuations around this small signal are large, and increase with $\ell$.
A way to see this is to introduce the random variable $u_\ell=(p_{n+\ell}-p_n).\varepsilon_n$.
By definition, ${\cal R}(\ell)$ is the average of $u_\ell$, and ${\cal D}(\ell)$ is 
the average
of $u_\ell^2$. Since ${\cal R}(\ell)$ is roughly constant whereas ${\cal D}(\ell)$ grows 
linearly 
with $\ell$, one sees that the impact of a given trade (as measured by ${\cal R}(\ell)$) 
rapidly becomes lost in the fluctuations. 

\begin{figure}
\begin{center}
\psfig{file=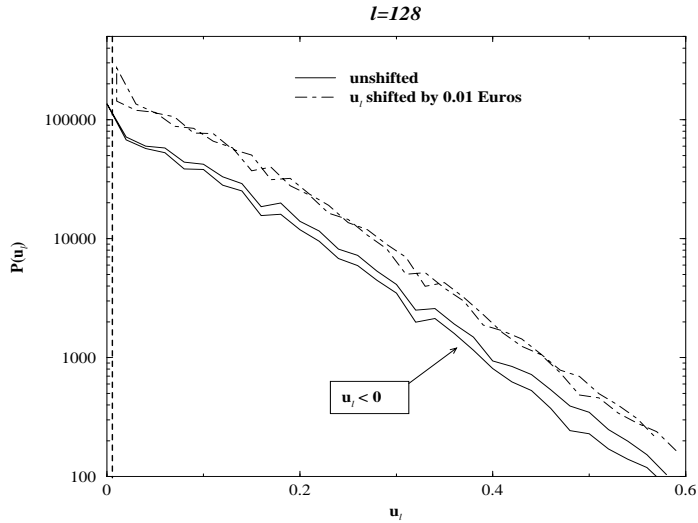,width=7cm,height=9.2cm,angle=270} 
\end{center}
\caption{Probability distribution $P(u_\ell)$ of the quantity 
$u_\ell=(p_{n+\ell}-p_n).\varepsilon_n$ (in Euros), for $\ell=128$. 
The data is again FT during 2002. 
The negative part of the distribution has been folded back to positive $u_\ell$ in order
to highlight the small positive skew of the distribution (which is seen to
increase slightly with $|u_\ell|$). The average value ${\cal R}(\ell) =
\langle u_\ell \rangle$ is shown as the vertical dashed line. The 
dashed-dotted line 
corresponds to the distribution of $u_\ell-\nu$ with $\nu=0.01$ Euros.
This curve has been shifted upwards for clarity.}
\label{Fig6}
\end{figure}

In Fig. 6, we show the whole empirical 
distribution $P(u_\ell)$
of $u_\ell$ for $\ell=128$ (but other values of $\ell$ lead to similar 
results). This distribution is found to be only slightly skewed 
in the direction of positive $u_\ell$. In fact, if one considers the shifted 
variable 
$u_\ell-\nu$, where $\nu=0.01$ Euros, the distribution becomes nearly 
symmetric. Note that $0.01$ Euros is equal to half the typical bid-ask spread 
and can therefore be seen as the cost of a market order. The Efficient Market picture suggests that the non zero value of $\langle u_\ell \rangle$ should 
mostly be due to a small fraction of informed trades, that correctly 
anticipate large price changes as a result of some private information, 
while most noise 
induced trades should only change the price on short time scales, before 
arbitrageurs set it back to its `true' value. 
In this case, the positive tail of the distribution $P(u_\ell)$ 
(corresponding to informed trades) should be much fatter than the 
negative tail. This 
asymmetry can in fact be taken as an objective measure of the fraction of 
informed trades. However, the nearly 
symmetric shape of $P(u_\ell-\nu)$ shown in Fig. 6 means that one can 
hardly detect the statistical presence of informed
trades that correctly anticipate the sign of the price change 
on a short term basis, such as to at least cover their trading costs.\footnote{Some of these trades might of course be profitable on the long run. But since 
the price process is nearly diffusive and that the number of buy and sell market orders are nearly equal, it is clear that difference between the fraction
of profitable trades on any given time scale and $50\%$ is small.}    
This result is consistent with the conclusion of 
other studies, where it is established that investors 
`trade too much'\cite{Odean}, and that the uninformed price pressure is large 
\cite{Hopman}. 
Note that $P(u_\ell)$ as defined above 
gives an equal weight to
all trades, independently of their volume. We have also considered 
the volume
weighted $P(u_\ell)$, which leads to the same qualitative conclusion.

The main conclusions of this section are thus that (a) large volumes 
impact prices on average much less 
(in relative terms) than smaller volumes, (b) the average impact of 
a given trade (as
measured by ${\cal R}(\ell)$) increases with time up to 
a certain time scale $\ell^*$ beyond 
which it decreases and (c) the fraction of trades that correctly anticipates
short term moves is small. 

\subsection{A Fluctuation-Response relation}

In the study of Brownian particles, a very important result that dates back to Einstein 
relates the diffusion coefficient $D$ to the response of the particle to an external force. 
That a similar relation might also hold in financial markets was first suggested by Rosenow 
\cite{Rosenow}, 
and substantiated there by some empirical results. We have performed an 
analysis related to, but different from that of Rosenow. For 
any given trading day, one can 
compute the average local diffusion constant ${\cal D}(\ell)$ over a 
given time scale, 
say $\ell=128$, and the average local price response ${\cal R}(\ell)$ 
over the same time scale. Rosenow, on the other hand, computes a
`susceptibility' as the slope of the average price change over a given 
time interval versus the volume imbalance during the same time interval 
(see \cite{Gopi}), and relates this susceptibility to the diffusion 
constant. The analogue of Rosenow's result \cite{Rosenow} 
(which was motivated by a 
Langevin equation for price variations -- see \cite{BC}), is a linear 
relation between ${\cal R}^2(\ell)$ and ${\cal D}(\ell)$, 
which we illustrate in 
Fig. 7 for FT, for two different periods (first semester of 2001, and 2002). A similar 
result can also be read from Fig. 3. As will be
clear in the following, such a relation will appear naturally within 
the simple model that we introduce in Section 3.

\begin{figure}
\begin{center}
\psfig{file=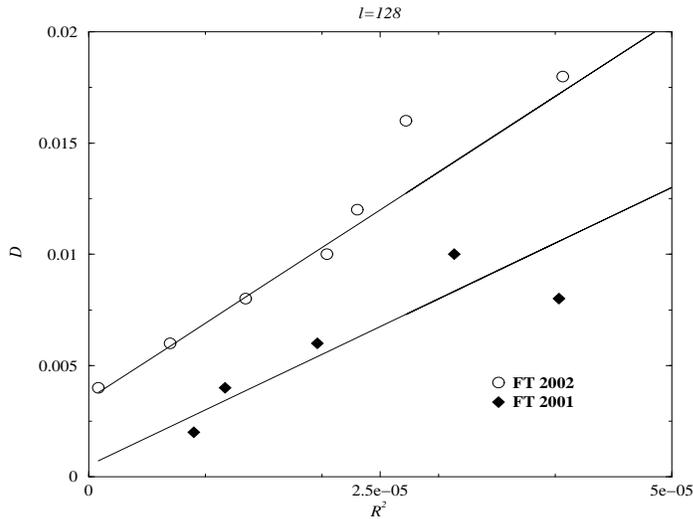,width=7cm,height=9.2cm,angle=270} 
\end{center}
\caption{Average diffusion constant $D={\cal D}(\ell)/\ell$, computed for $\ell=128$, and
conditioned to a certain value of ${\cal R}^2(\ell)$, also computed for $\ell=128$ (FT). The open
symbols correspond to 2002, whereas the black symbols are computed using the first semester 
of 2001, where the tick size was 5 times larger. Correspondingly, the $x$-axis was rescaled down
by a factor $25$ and the $y$-axis by a factor five for this data set.}
\label{Fig7}
\end{figure}

\subsection{Long term correlation of trade signs}

All the above results are compatible with a `zero intelligence' picture of financial markets,
where each trade is random in sign and shifts the price permanently, because all other 
participants update their evaluation of the stock price as a function of the last trade. 
As shown in \cite{Maslov,Challet1,Farmer1,BMP,Farmer2,Challet2}, a 
model of the order book based on a purely random order flow indeed allows one to go quite far 
in the quantitative understanding of financial markets. 
In this context, the concave shape of
the impact as a function of the volume can be understood as an 
order book effect, where the
average size of the queue increases with depth.\footnote{However, other 
effects are probably important to understand this concavity, such as the 
conditioning of large market orders to the size of the order book -- see 
\cite{Bali,Hopman}.}  

This model of a totally random stock market
is however qualitatively incorrect for the following reason. Although, as mentioned above, the 
statistics of price changes reveals very little temporal correlations, the correlation
function of the sign $\varepsilon_n$ of the trades, on the other hand, reveals very slowly
decaying correlations. This correlation has been mentioned in some papers 
before, see e.g. \cite{Hopman}. Here, we propose that these correlations
decay as a power-law of the time lag. 
 
More precisely, one can consider the following correlation function:
\be
{\cal C}_0(\ell)= \langle \varepsilon_{n+\ell}\varepsilon_n \rangle -
\langle \varepsilon_n \rangle^2
\ee
If trades were random, one should observe that ${\cal C}_0(\ell)$ decays to zero
beyond a few trades. Surprisingly, this is not what happens: on the contrary, ${\cal C}_0(\ell)$ is strong and 
decays very slowly toward zero, as an inverse power-law of $\ell$ (see Fig. 8):
\be
{\cal C}_0(\ell) \simeq \frac{C_0}{\ell^\gamma}, \qquad (\ell \geq 1).
\ee
The value of $\gamma$ seems to be somewhat stock dependent. 
For example, for FT, one finds 
$\gamma \approx 1/5$, whereas for Total $\gamma \approx 2/3$. In their 
study, Lillo and Farmer found a somewhat larger value of 
$\gamma \approx 1/2$ for Vodafone \cite{Farmer3}. In any case, the value of 
$\gamma$ is 
found to be smaller than one, which is very important because the integral of ${\cal C}_0(\ell)$
is then {\it divergent}. Now, as will be shown more precisely in the next section, the integral of ${\cal C}_0(\ell)$
can intuitively be thought of as the effective number $N_e$ of correlated successive trades.
Hence, 
out of -- say -- 1000 trades, one should group together 
\be
N_e \simeq 1+ \sum_{\ell=1}^{1000} {\cal C}_0(\ell) \approx 1 + \frac{C_0}{1-\gamma} 
1000^{1-\gamma}
\ee
`coherent' trades. For FT, $\gamma \approx 1/5$ and $C_0 \approx 0.2$, which means that the
effect of one trade should be amplified, through the correlations, by a factor $N_e \approx 50$ !
In other words, both the response function ${\cal R}$ and the diffusion constant should
increase by a factor $50$ between $\ell=1$ and $\ell=1000$, in stark contrast with the 
observed empirical data. This is the main puzzle that one should try 
to elucidate: how can
one reconcile the strong, slowly decaying correlations 
in the sign of the trades with the 
nearly diffusive nature of the price fluctuations, and the nearly structureless response function?

\begin{figure}
\begin{center}
\psfig{file=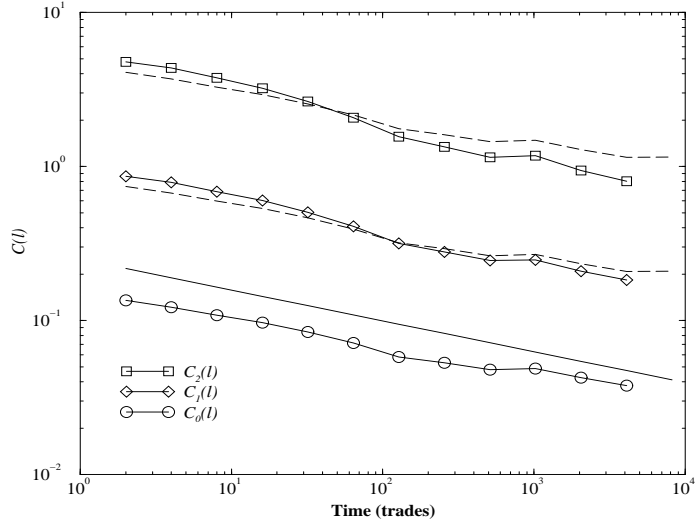,width=7cm,height=9.2cm,angle=270} 
\end{center}
\caption{Volume weighted sign autocorrelation functions as a function of time lag: ${\cal C}_0$,
${\cal C}_1$, ${\cal C}_2$ (see text for definitions). The straight line corresponds to 
$\ell^{-\gamma}$ with $\gamma=1/5$. The dotted lines correspond to the simple 
approximation given by Eqs. (\protect\ref{correlprop}).}
\label{Fig8}
\end{figure}
Before presenting a mathematical transcription of the above question and proposing a possible 
resolution, let us comment on two related correlation functions that will naturally 
appear in the following, namely:
\be
{\cal C}_1(\ell)= \langle \varepsilon_{n+\ell}\,\, \varepsilon_n \ln V_n \rangle,
\ee
and
\be
{\cal C}_2(\ell)= \langle \varepsilon_{n+\ell} \ln V_{n+\ell}\,\, \varepsilon_n \ln V_n \rangle.
\ee
We have found empirically that these two `mixed' correlation functions are proportional 
to ${\cal C}_0(\ell)$ (see Fig 8):
\be\label{correlprop}
{\cal C}_1(\ell) \approx \langle \ln V \rangle {\cal C}_0(\ell); \qquad 
{\cal C}_2(\ell) \approx \langle \ln V \rangle^2 {\cal C}_0(\ell).
\ee
There are however small systematic deviations, which indicate that (i) small
volumes contribute more to the long range correlations that larger volumes
and (ii) $\ln V - \langle \ln V \rangle$ is a quantity exhibiting long 
range correlations as well.

\section{A micro-model of price fluctuations}

\subsection{Set up of the model}

In order to understand the above results, we will 
postulate the following trade 
superposition model, where the price at time $n$ is written as a sum over all past 
trades, of the impact of one given trade propagated up to time $n$:
\be\label{model}
p_n = \sum_{n'<n} G_0(n-n') \varepsilon_{n'} \ln V_{n'} + 
\sum_{n'<n} \eta_{n'},
\ee
where $G_0(.)$ is the `bare' impact function (or propagator) of a single trade, that
we assume to be a fixed, non random function that only depends on time differences. 
The $\eta_n$ are also random variables, assumed to be independent from the 
$\varepsilon_n$ and model all sources of price changes not 
described by the direct
impact of the trades: the bid-ask can change as the result of some news, or
of some order flow, in the absence of any trades. We will in the following assume that the $\eta_n$ are also uncorrelated in time, although this assumption
can easily be relaxed.
In the above model, we
assume that the `bare' impact function $G_0$ is not itself fluctuating, 
which can only be an approximation.

The bare impact function $G_0(\ell)$ represents by definition the 
average impact of a single 
trade after $\ell$ trades. It could be in principle measured empirically by 
launching on the market a
sequence of real trades of totally random signs, and averaging the impact over this sample
of trades (a potentially costly experiment!).\footnote{However, following this
procedure might induce `copy-cat' trades and still lead to a 
difference between the measured response function and $G_0$} 
As will be clear below, the difference between the quantity ${\cal R}(\ell)$
introduced in the previous Section and $G_0(\ell)$ in fact 
comes from the strong 
autocorrelation of the sign of the trades. In order to understand 
the temporal structure 
of $G_0(\ell)$, note that a single trade first impacts the midpoint by changing the bid 
(or the ask). But then the subsequent limit order flow due to that particular trade might 
either center on average around the new midpoint (in which case $G_0(\ell)$ would be constant), or, as
we will argue below, tend to mean revert toward the previous midpoint 
(in which case $G_0(\ell)$ decays with $\ell$). As discussed below (see
Eq. (\ref{gains})), 
the asymptotic behaviour of the bare impact function in fact 
reveals the average cost of a single market order: if $G_0(\ell \gg 1)/G_0(1)$ 
is small, the cost is large since the initial impact of the trade  
is only temporary, and is not followed by a true long term change of the 
price. 

Using this representation, the price increment between an arbitrarily chosen initial 
time $0$ and time $\ell$ is:
\be
p_{\ell}-p_0 = \sum_{0 \leq n < \ell} G_0(\ell-n) \varepsilon_{n} \ln V_{n} +
\sum_{n < 0} \left[G_0(\ell-n)-G_0(-n)\right] \varepsilon_{n} \ln V_{n}
+\sum_{0 \leq n < \ell} \eta_{n} .
\ee
If the signs $\varepsilon_{n}$ were independent random variables, both the response 
function and the diffusion would be very easy to compute. For example, one would have:\footnote{In the following, we will use the subscript `t' to denote the theoretical
expressions for the response function or diffusion.}
\be
{\cal R}_t(\ell) = \langle \ln V \rangle G_0(\ell),
\ee
i.e. the observed impact function and the bare response function would be proportional.
Similarly, one would have:
\be
{\cal D}_t(\ell) = \langle \ln^2 V \rangle \left(\sum_{0 < n \leq \ell} G_0^2(n) +
\sum_{n > 0} \left[G_0(\ell+n)-G_0(n)\right]^2 \right) + D_\eta \ell,
\ee
where $D_\eta$ is the variance of the $\eta$'s.
In the simplest case of a constant bare impact function, $G_0(\ell)=\Gamma_0$ for all $\ell > 0$, one then 
finds a pure diffusive behaviour, as expected:
\be
{\cal D}_t(\ell) = \ell \left[\langle \ln^2 V \rangle \Gamma_0^2 + D_\eta\right]. 
\ee
This result (no correlations between the $\varepsilon$'s and a constant bare impact function)
corresponds to the simplest possible zero intelligence market. However, we have seen that 
in fact the $\varepsilon$'s have long range correlations. In this case, the average response 
function reads: 
\be \label{response}
{\cal R}_t(\ell) = \langle \ln V \rangle G_0(\ell)+ 
\sum_{0 < n < \ell} G_0(\ell-n) {\cal C}_1(n) + 
\sum_{n > 0} \left[G_0(\ell+n)-G_0(n)\right] {\cal C}_1(n).
\ee
Note in passing that our trade superposition model, Eq. (\ref{model}), together with 
Eq. (\ref{correlprop}) leads to the factorization property mentioned above (see Fig.  5):
\be
{\cal R}_t(\ell,V)= \frac{\ln V}{\langle \ln V \rangle} {\cal R}_t(\ell).
\ee
Now, one sees more formally the paradox discussed in the previous Section: assuming that
the impact of each trade is permanent, i.e. $G_0(\ell)=\Gamma_0$, leads to:
\be
{\cal R}_t(\ell) = \Gamma_0 \left[\langle \ln V \rangle + \sum_{0 < n < \ell} {\cal C}_1(n)\right].
\ee
If ${\cal C}_1(n)$ decays as a power-law with an exponent $\gamma < 1$, 
then the average 
impact ${\cal R}(\ell)$ should {\it grow} like $\ell^{1-\gamma}$, and therefore be amplified 
by a very large factor as $\ell$ increases, at variance with empirical data. 
The only way out of this conundrum is (within
the proposed model) that the bare impact function $G_0(\ell)$ itself should decay with time, in 
such a way to offset the amplification effect due to the trade correlations. 

\subsection{A relation between the bare propagator and the sign correlation function}

In order to get some guidance, let us now look at the 
general formula for the diffusion. 
After a few lines of calculations, one finds:
\bea\label{diff}
{\cal D}_t(\ell) &=& \langle \ln^2 V \rangle \left[\sum_{0 \leq n < \ell} G_0^2(\ell-n)
+ \sum_{n > 0} \left[G_0(\ell+n)-G_0(n)\right]^2\right] \\
&+& 2\Delta(\ell)+ D_\eta \ell,
\eea
where $\Delta(\ell)$ is the correlation induced contribution:
\bea
\Delta(\ell) &=& \sum_{0 \leq n < n' < \ell} G_0(\ell-n) G_0(\ell-n') {\cal C}_2(n'-n)\\
\nonumber
&+& \sum_{0 < n < n'} \left[G_0(\ell+n)-G_0(n)\right]\left[G_0(\ell+n')-G_0(n')\right]
{\cal C}_2(n'-n) \\
&+& \sum_{0 \leq n < \ell} \sum_{n'> 0} G_0(\ell-n)
\left[G_0(\ell+n')-G_0(n')\right]{\cal C}_2(n'+n).
\eea

The constraint from empirical data is that this expression must be 
approximately
linear in $\ell$. As shown in the Appendix, the requirement that 
${\cal D}_t(\ell)$ is strictly linear in $\ell$ for all $\ell$ in fact allows one 
to express $G_0(\ell)$ as a function of ${\cal C}_2(\ell)$. Here, we 
present a simple asymptotic argument. If we make the 
ansatz that the bare impact 
function $G_0(\ell)$ 
also decays as a power-law:
\be
G_0(\ell)=\frac{\Gamma_0 \ell_0^\beta}{(\ell_0+\ell)^\beta} \qquad (\ell \geq 1)
\ee
then one can estimate ${\cal D}_t(\ell)$ in the large $\ell$ limit. 
When $\gamma < 1$,
one again finds that the correlation induced term $\Delta(\ell)$ is dominant, and
all three terms scale a $\ell^{2-2\beta-\gamma}$, provided $\beta < 1$. In other words, the
Hurst exponent of price changes is given by $2H=2-2\beta-\gamma$. Therefore, the condition 
that the 
fluctuations are diffusive at long times ($H=1/2$) imposes a relation 
between the decay 
of the sign autocorrelation $\gamma$ and the decay of the bare impact function $\beta$ 
that reads:
\be
2 \beta + \gamma = 1 \longrightarrow \beta_c=\frac{1-\gamma}{2}
\ee 
For $\beta > \beta_c$, the price is {\it sub-diffusive} ($H < 1/2$), which means that price 
changes show anti-persistence; 
while for $\beta < \beta_c$, the price is {\it super-diffusive} ($H > 1/2$), i.e. price changes 
are persistent. 
For FT, $\gamma \approx 1/5$ and therefore $\beta_c \approx 2/5$.

As shown in the Appendix, one can in fact obtain an exact relation between
$G_0(\ell)$ and ${\cal C}_2(\ell)$ if one assumes that price changes are
strictly uncorrelated (i.e. that ${\cal D}(\ell)$ is linear in $\ell$ for all 
$\ell$). The asymptotic analysis of this relation leads, not surprisingly, to
the same exponent relation $\beta_c=({1-\gamma})/{2}$ as above. 

At this stage, there seems still to be a contradiction with empirical data, 
for if one goes back to the response function given by Eq. (\ref{response}), 
one finds that whenever $\beta + \gamma < 1$ 
(which is indeed the case for $\beta=\beta_c$ and $\gamma < 1$), the
dominant contribution to ${\cal R}_t(\ell)$ should behave as 
$\ell^{1-\beta-\gamma}$ 
and thus {\it grow} with $\ell$. For example, for $\gamma \approx 1/5$ and $\beta \approx 2/5$,
one should find that ${\cal R}_t(\ell) \propto \ell^{2/5}$, which is incompatible with 
the empirical data of Figs. 3 and 4. But the surprise comes from
the numerical prefactor 
of this power law. One finds, for large $\ell$:
\be
{\cal R}_t(\ell) \simeq \langle \ln V \rangle \Gamma_0 C_0 \, \frac{\Gamma(1-\gamma)}{\Gamma(\beta)
\Gamma(2-\beta-\gamma)} \left[\frac{\pi}{\sin \pi \beta} - 
\frac{\pi}{\sin \pi (1-\beta-\gamma)}\right] \ell^{1-\beta-\gamma}.
\ee
Therefore, only when $\beta = \beta_c$, is the prefactor exactly zero, 
and leads to the
possibility of a nearly constant impact function! For 
faster decaying impact functions (larger $\beta$'s), this prefactor is 
negative, 
whereas for more slowly decaying impact functions this prefactor is 
positive.\footnote{Note that although this prefactor increases (in absolute value) with 
$\beta$ for $\beta > \beta_c$, the power of $\ell$ decreases, which means that
for large $\ell$ the amplitude of ${\cal R}_t(\ell)$ decreases with $\beta$, as
intuitively expected.} Interestingly, even if the bare response function $G_0(\ell)$ is
positive for all $\ell$, the average response ${\cal R}_t(\ell)$ can 
become negative for
large enough $\beta$'s, as a consequence of the correlations between trades.

\subsection{Fitting the average response function}

Since the dominant term is zero for the `critical' case 
$\beta = \beta_c$, and
since we are interested in the whole function ${\cal R}_t(\ell)$ 
(including the small $\ell$
regime), we have computed ${\cal R}_t(\ell)$ numerically, by performing 
the discrete sum Eq. (\ref{response}) exactly, and fitted it to the
empirical response ${\cal R}$. The results are shown in Fig. 9. We have 
fixed the parameters $\gamma$ and $C_0$ to the values extracted from the
behaviour of ${\cal C}_1(\ell)$ (see Fig. 8): $\gamma=0.24$ and $C_0=0.20$.    
The overall scaling parameter $\Gamma_0$ is adjusted to 
$\Gamma_0=2.8\, 10^{-3}$ Euros to match the value of ${\cal R}(\ell=1)$. 
The values of $\beta$ and $\ell_0$ are fitting parameters: we show in Fig. 9
the response function computed for different values of $\beta$ in the vicinity 
of $\beta_c=0.38$, and used $\ell_0=20$. 

The results are compared with the empirical data for FT, showing that one can
indeed satisfactorily reproduce, when $\beta \approx \beta_c$, 
a weakly increasing impact function
that reaches a maximum and then decays. One also sees, from Fig. 9, 
that the relation
between $\beta$ and $\gamma$ must be quite accurately satisfied, otherwise the 
response function shows a distinct upward trend (for $\beta < \beta_c$) or a 
downward trend ($\beta > \beta_c$).\footnote{This might actually explain the
different behaviour of Pechiney seen in Fig. 4.} 
In fact, we have tried other simple forms for 
$G_0(\ell)$, such as a simple exponential decay toward a possibly non zero asymptotic value, 
but this leads to unacceptable shapes for ${\cal R}(\ell)$.

\begin{figure}
\begin{center}
\psfig{file=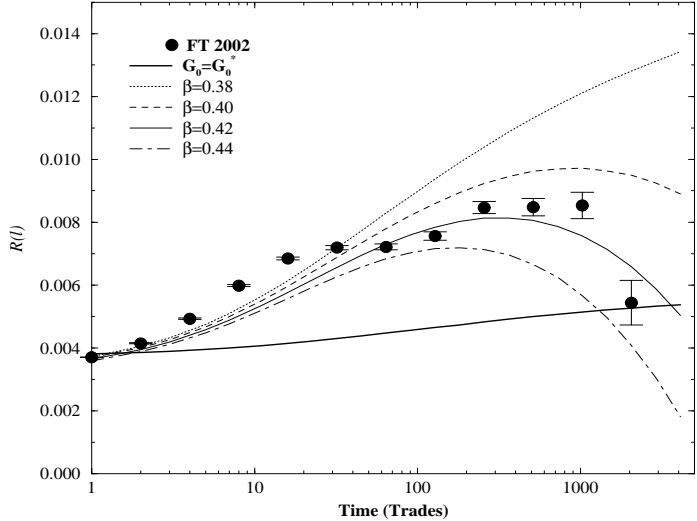,width=7cm,height=9.2cm,angle=270} 
\end{center}
\caption{Theoretical impact function ${\cal R}_t(\ell)$, 
from Eq. (\ref{response}), and for 
different values of $\beta$ close to $\beta_c=0.38$. 
The shape of the empirical response
function can be quite accurately reproduced using $\beta=0.42$. The only 
remaining free parameter is $\ell_0 = 20$. The thick plain line is 
${\cal R}_t(\ell)$ computed using the `pure diffusion' propagator 
$G_0^*$ determined in Appendix, 
Eq. (\ref{exact}).}
\label{Fig9}
\end{figure}

It is also interesting to use the propagator $G_0^*$ determined in the 
Appendix from the assumption of a purely diffusive price process for 
all $\ell$'s.  This propagator is plotted in Fig. 10, and compared to 
the $G_0$ determined above from the fit of ${\cal R}(\ell)$. As shown in
Fig. 9, the use of $G_0^*$ does {\it not} lead to a very good fit 
of ${\cal R}(\ell)$. Since the latter quantity is in fact {\it very sensitive 
to the chosen shape} for $G_0$, it does reveal small, 
but systematic deviations from a purely diffusive price process. [Note that
if one had ${\cal C}_2(\ell)={\cal C}_1(\ell)$, the resulting ${\cal R}(\ell)$
should be strictly constant.] 

\begin{figure}
\begin{center}
\psfig{file=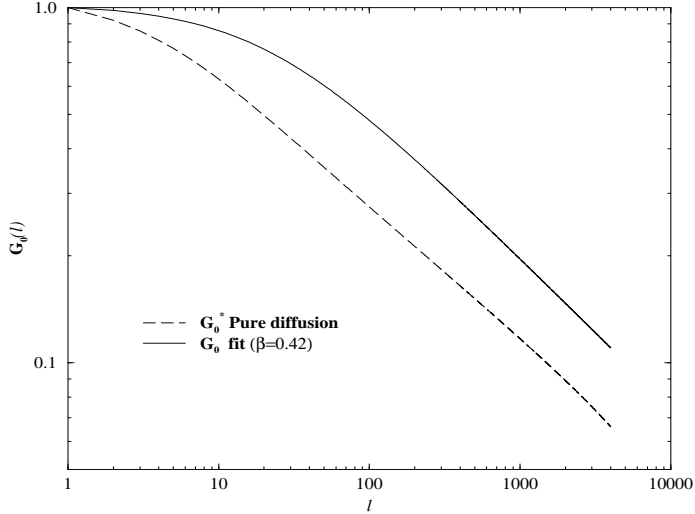,width=7cm,height=9.2cm,angle=270} 
\end{center}
\caption{Shape of the bare propagator $G_0$, determined either 
by the fit of $\cal R$, with $\beta=0.42$ and $\ell_0=20$, or using the 
exact relation, Eq. (\ref{exact}), derived in the Appendix from the 
assumption of a purely diffusive process.}
\label{Fig11}
\end{figure}

\subsection{Back to the diffusion constant}

As we showed above, the reason for the fine tuning of $\beta$ is the 
requirement that price 
changes are almost diffusive. We can therefore also compute ${\cal D}_t(\ell)$ 
for all values of $\ell$ using the very same values of 
$\gamma$, $\beta$, $C_0$, $\ell_0$ and $\Gamma_0$. Now,
in order to fit the data one has two extra free parameters: one is $D_\eta$, 
and the other comes about because the mid-point can change without any trade. 
One should thus add to ${\cal D}_t(\ell)$ an $\ell$-independent `error' term 
$D_0$ that survives in the $\ell=0$ limit, and is associated to 
bid-ask fluctuations. 
With these two extra parameters, one can 
reproduce the empirical determination of ${\cal D}(\ell)/\ell$ 
(see Fig. 11). The small deviations of this quantity from a horizontal line at
finite $\ell$ are due to the difference between 
$G_0$ and $G_0^*$ and/or to the possible autocorrelations between the 
$\eta_n$ variables, which we have neglected here. 
Note that the contribution of 
the term $D_\eta$ turns out to be a factor two larger than that of
the impact contribution, Eq. (\ref{diff}), which means that the small increase 
of the `impact contribution' with $\ell$ (lower graph of Fig. 11) is hardly 
detectable in ${\cal D}(\ell)/\ell$.

\begin{figure}
\begin{center}
\psfig{file=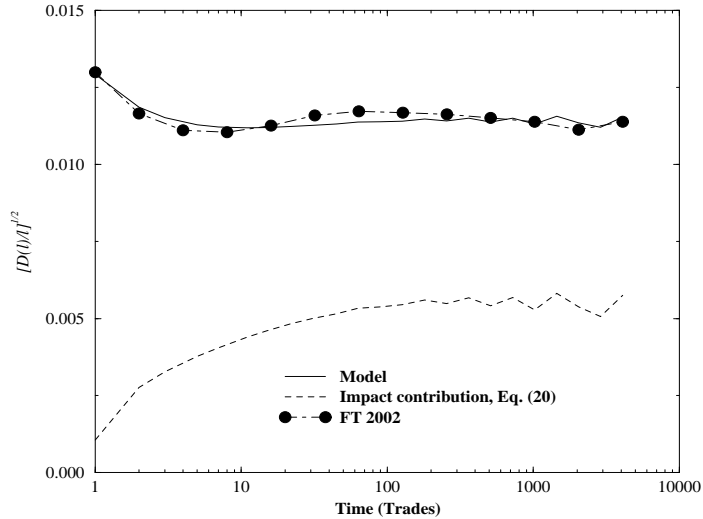,width=7cm,height=9.2cm,angle=270} 
\end{center}
\caption{Diffusion constant ${\cal D}(\ell)/\ell$, using Eq. (\ref{diff}), 
with the values of $\gamma$, $\beta$, $C_0$, $\ell_0$ and $\Gamma_0$ determined from ${\cal R}(\ell)$. 
Two extra parameters were 
used: $D_\eta = 10^{-4}$ and $D_0=6.6 \, 10^{-5}$ (both in Euro squared).
The lower graph is the `impact contribution' to ${\cal D}_t(\ell)$, 
given by Eq. 
(20) with $D_\eta=0$. The `oscillations' at long times is a 
numerical artefact.}
\label{Fig10}
\end{figure}

Coming back to the Fluctuation-Response relation discussed in Section 2.4, we
see that our model predicts, for $\ell \gg 1$ where the effect of $D_0$ can be
neglected: 
\be
\frac{{\cal D}_t(\ell)}{\ell} = Z \langle \ln V \rangle^2 C_0 \Gamma_0^2 + D_\eta
,\qquad  {\cal R}_t(\ell) = Z'\langle \ln V \rangle \Gamma_0 C_0,
\ee
where $Z,Z'$ are numerical constants.
Assuming that from one day to the next both the average (log-)traded volume and the 
impact $\Gamma_0$ of each individual trade might change, while $C_0$ is fixed, immediately leads 
to the affine relation between ${\cal D}$ and ${\cal R}^2$ reported in Section 2.4.

\subsection{Discussion}

The conclusion of this Section is that our `micro-model' of prices, 
Eq. (\ref{model}), can be used as 
a theoretical canvas to rationalize and interpret the empirical results found in the 
previous Section.
Most surprising is the constraint that the empirical results impose on the shape of the `bare' 
response function $G_0$, which is found to be a slowly decaying power law which must 
precisely cancel the slowly decaying autocorrelation of the trades, but reveals 
systematic deviations from a pure diffusion process, hardly noticeable on the
diffusion constant itself. The fact that the bare impact 
function decays with time (at least on intra-day time scales), 
in a finely tuned way to compensate the long memory in the trades, 
is the central result of this paper. This effect is lost in the
zero intelligence models of Poisonnian order flows, where, after decreasing during 
a short transient, the impact of each trade becomes permanent: 
$G_0(\ell) \to G_\infty >0$.
(On this point, see the model studied in \cite{Farmer2}, where it is shown 
that prices are sub-diffusive on time scales shorter than the life-time 
of limit orders, essentially as a consequence of the shape of the order book). 
In fact, both the long time memory of the trades and the slowly relaxing impact
function reported here must be the consequence of the strategic behaviour of market 
participants, that 
we discuss below in order to get an intuitive understanding of the mechanisms at play.

Although our detailed analysis concerns FT, it is clear that our conclusions are more general,
since both the strong autocorrelations in the trade signs, the near constancy of the 
average response function and the diffusive nature of price changes have been observed on all
stocks, with only quantitative changes (see Figs 2 and 4). It would be interesting to 
document these 
quantitative difference, and relate these to liquidity, or to the size of the bid-ask spread.

Finally, it would be very interesting
to know whether the bare response function levels off to a finite value for 
large time lags; this will require 
to go beyond the analysis of the present paper and to deal with overnight 
effects to enlarge the available 
range of $\ell$ values. However, it seems reasonable to expect 
that $G_0(\ell)$ should indeed reach a finite 
asymptotic value for values of $\ell$ corresponding to a few days of 
trading.\footnote{Hopman quotes three days as the time beyond which the
autocorrelation of the trades sign falls to zero \cite{Hopman}.} 

\section{Critical balance of opposite forces: Market orders vs. limit orders}

Although trading occurs for a large variety of reasons, it is useful to 
recognize that traders organize in two broad categories: 
\begin{itemize} 
\item One is that of `liquidity takers', that trigger trades by 
putting in market orders. The motivation for this category of traders might 
be to take advantage of some `information', and 
make a profit from correctly anticipating future price changes. Information 
can in fact be of very different nature: fundamental (firm based), 
macro-economical,  
political, statistical (based on regularities of price patterns), etc. 
Unfortunately, information is often hard to interpret correctly, and it is 
probable that many of these `information' driven trades are misguided (on this
point, see the remarkable work of Odean \cite{Odean}; see also \cite{Hopman}
and refs. therein). 
For example, systematic hedge funds which
take decisions based on statistical pattern recognition have a typical success 
rate of only $52 \%$. There is no compelling reason to believe that the 
intuition of traders in markets room fares much better than that. 
Since market orders
allows one to be immediately executed, many impatient investors, who want to 
liquidate their position, or hedge, etc. might be tempted to place 
market orders, 
even at the expense of the bid-ask spread $s(t)=a(t)-b(t)$. 

\item The other category is that of `liquidity
providers' (or `market makers', although on 
electronic markets all participants can act as liquidity providers 
by putting in limit orders), who offer to buy or to sell but 
avoid taking any bare position on the market. Their profit comes 
from the bid-ask spread $s$: the sell price is always slightly larger 
than the buy price, so that each round turn 
operation leads to a profit equal to the spread $s$, at least 
if the midpoint has not changed in the mean time (see below).
\end{itemize}

This is where the game becomes interesting. Assume that a liquidity taker
wants to buy, so that an increased number of buy orders arrive on the market. 
The liquidity providers is tempted to increase
the offer (or ask) price $a$ because the buyer might be informed and really 
know 
that the current price is too low and that it will most probably increase in 
the near future. Should this happen, the liquidity provider, who has to close 
his position later, might 
have to buy back at a much higher price and experience a loss. 
In order not to trigger a sudden increase of $a$
that would make their trade costly, liquidity takers obviously 
need to put on not
too large orders. This is the rationale for 
dividing one's order in small chunks and disperse these as much as possible 
over time so as not to appear on the `radar screens'. 
Doing so liquidity
takers {\it necessarily create some temporal correlations} in the sign of 
the trades. Since these traders probably have a somewhat broad spectrum of 
volumes to trade \cite{NatureG}, and therefore of trading horizons (from a few 
minutes to several weeks), this can easily explain the slow, 
power-law decay of 
the sign correlation function ${\cal C}_0(\ell)$ reported above. 

Now, if the market orders in fact do {\it not} contain useful information 
but are the result 
of hedging, noise trading, misguided interpretations, errors, etc., 
then the price should not 
move up on the long run, and should eventually mean revert to its previous 
value. Liquidity providers
are obviously the active force behind this mean reversion, again because 
closing their position will be
costly if the price has moved up too far from the initial price. More
precisely, a computation of the 
liquidity provider average gain per share $\cal G$ can be performed 
\cite{ustocome}, 
and is found to be, for trades of volume $V$:
\be\label{gains}
{\cal G} = s + {\cal R}(0,V) - {\cal R}(\infty,V) \approx s + 
\ln V \left[{\cal R}(0) - {\cal R}(\infty)\right],
\ee
where ${\cal R}(0,V)$ is the immediate average impact of a trade, 
before new limit orders 
set in. We have in fact checked empirically that ${\cal R}(0,V) \approx {\cal R}(1,V)$. 
From the above
formula, one sees that {\it it is in the interest of liquidity 
providers to mean revert the price}, such as to make ${\cal R}(\infty)$ 
as small as possible. However, this mean reversion 
cannot take place too quickly, again because 
a really informed trader would then be able to buy a large volume at a 
modest price. Hence, this mean reversion must be slow. 
From the quantitative analysis of 
Section 3, we have found that there is hardly any
mean reversion at all on short time scales $\ell < \ell_0$, and that this 
effect can be described as a slow power-law for larger $\ell$'s.  
Actually, the action of liquidity providers and liquidity takers must be 
such that no (or very little) 
linear correlation is created in the price changes, 
otherwise statistical arbitrage opportunities would be created at the 
detriment of one or the other population. 

To summarize: liquidity takers must dilute their orders and create long 
range correlations in the trade 
signs, whereas liquidity providers must correctly handle the fact 
that liquidity takers might either possess useful information 
(a rare situation, but that can be very costly since the 
price can jump as a result of some significant news), or might not be 
informed at all and trade randomly. 
By slowly mean reverting the price, market makers 
minimize the probability 
that they either sell too low, or have to buy back too high. The 
delicate balance 
between these 
conflicting tendencies conspire to put the market at the border 
between persistence (if mean reversion is
too weak, i.e. $\beta < \beta_c$) or 
anti-persistence (if mean reversion is too strong, i.e. $\beta > \beta_c$), 
and therefore eliminate arbitrage opportunities.

It is actually enlightening to propose a simple model that could explain 
how market makers enforce this
mean reversion.\footnote{We have in fact directly checked on the data that 
the evolution 
of the midpoint between trades (resulting from the order flow) is indeed 
anticorrelated with the impact of the trades. On this point, see also
\cite{Hopman} where the limit order flow subsequent to a trade is studied.} 
Assume that upon placing limit orders, there is a 
systematic bias toward some moving 
average of past prices. If this average is for simplicity taken to be an 
exponential 
moving average, the
continuous time description of this will read:
\bea\label{meanrevert}
\frac{dp_t}{dt} &=& - \Omega(p_t - \overline{p}_t) + \eta_t \\
\frac{d\overline{p}_t}{dt} &=&  \kappa (p_t - \overline{p}_t),
\eea   
where $\eta_t$ is the random driving force due to trading, $\Omega$ the 
inverse time 
scale for 
the strength of the mean reversion, and $1/\kappa$ the `memory' time over 
which the average 
price $\overline{p}_t$ is computed. The first equation means that liquidity 
providers tend to mean revert the price toward $\overline{p}_t$, while the 
second describes the update of the exponential moving average $\overline{p}_t$ with 
time. This set of linear equations can be solved, and leads to a solution of the
form $p_t =\int^t dt' G_0(t-t') \eta_t'$, with a bare propagator given by:
\be
G_0(t) = 
(1-G_\infty) \exp[-(\Omega+\kappa)t]+G_\infty,
\ee 
i.e. an exponential decay toward a finite 
asymptotic value $G_\infty=\kappa/(\Omega+\kappa)$. Note that, interestingly,
it is the self-referential effect that leads to a non zero asymptotic impact. 
If the fundamental price was known to all, $\kappa=0$ and $G_\infty=0$. In the opposite limit 
where $\kappa \gg \Omega$, the last price is taken as the reference price, and 
$G_\infty \to 1$.
A way to obtain $G_0(t)$ to resemble a power-law is to 
assume that 
different market makers
use different time horizons to compute a reasonable reference price. 
This leads to a $G_0(t)$ which 
writes as the sum of time exponentials with different rates 
which can easily mimic a pure power-law. 

The message of the above model is actually quite interesting from the point 
of view of Efficient Markets: it suggests that nobody 
really knows what the correct 
reference price should be, and that its best
proxy is in fact its own past average over some time window 
(the length of which being itself distributed over several time scales). 

\section{Summary and Conclusion}

The aim of this paper was to study in details the statistics of price changes at the 
trade by trade level, and to analyze the interplay between the impact of each trade 
on the price and the volatility. Empirical data shows that (a) the price (midpoint) process
is close to being purely diffusive, even at the trade by trade scale (b) the temporal structure 
of the impact function first increases and reaches a maximum after $100-1000$ trades, before
decreasing back, with a rather limited overall variation (typically a factor 2) and (c) 
the sign of the trades shows surprisingly long range (power-law) correlations. The paradox
is that if the impact of each trade was permanent, 
the price process should be strongly super-diffusive and the
average response function should increase by a large factor as a 
function of the time-lag. 

As a possible resolution of this paradox, we have proposed a micro-model of prices, 
Eq. (\ref{model}) 
where the price at any instant is the causal result of all past trades, mediated by what we
called a bare impact function, or propagator $G_0$. All the empirical results can be reconciled 
if one assumes that this bare propagator also decays as a power-law in time, with an exponent 
which is precisely tuned to a critical value, ensuring simultaneously that prices are 
diffusive on long time scales and that the response function is 
nearly constant. Therefore, the seemingly 
trivial random walk behaviour of price changes in fact results from a 
fined-tuned competition
between two opposite effects, one leading to super-diffusion 
(the autocorrelation of trades) 
and the other leading to sub-diffusion 
(the decay of the bare impact function). The
cancellation is however not exact: the non trivial 
behaviour of the average response
function allows one to detect small, but systematic 
deviations from a purely diffusive
behaviour, deviations that are hardly detectable on 
the price fluctuations themselves.

In financial
terms, the competition is between liquidity takers, 
that create long range correlations by 
dividing their trading volume in small quantities, and 
liquidity providers that tend to 
mean revert the price such as to optimize their gains (see Eq. (\ref{gains})).
The resulting absence of correlations in price changes, and therefore of arbitrage
opportunities is often postulated {\it a priori} in the economics literature, 
but the details of the mechanism that removes these arbitrage 
opportunities are rather obscure. The main message of this paper is that 
the random walk nature of price changes is not due to the unpredictable 
nature of incoming news, but appears as a dynamical consequence of the 
competition between antagonist market forces. In fact, the role of real (and
correctly interpreted) news appears to be rather thin: we have defined a model independent indicator 
of the fraction of `informed' trades, as the asymmetry 
of the probability distribution of the {\it signed} price variation, 
where the sign is that of 
the trade at the initial time. Information triggered trades should reveal 
in a detectable
positive skew of this distribution, in particular in the tails. Consistently 
with other studies \cite{Odean}, our 
empirical results only show very weak asymmetry, barely sufficient to cover trading costs, 
which means that only 
a small fraction of trades can a posteriori 
described as truly informed, whereas most trades can be classified as noise. 
This result is most probably one of the mechanism needed to explain 
the excess volatility puzzle first raised by 
Schiller \cite{Schiller}.

From a more general standpoint, our finding that the absence of arbitrage 
opportunities results from a critical balance 
between antagonist effects is quite interesting. It 
might justify several claims made in the (econo-)physics literature that 
the anomalies in price
statistics (fat tails in returns described by power laws \cite{Lux,Stanley}, 
long range 
self similar volatility correlations \cite{Cont,Muzy}, and the long ranged 
correlations in signs reported here and in \cite{Farmer3}) are due to 
the presence of a critical 
point in the vicinity of which the market operates 
(see e.g. \cite{soc}, and in the
context of financial markets \cite{MG,QF}). If a fine-tuned
balance between two competing effects is needed to ensure 
absence of arbitrage opportunities, one should expect that 
fluctuations are crucial, since 
a local unbalance 
between the competing forces can lead to an instability. In this respect, 
the
analogy with the balancing of a long stick is quite enticing \cite{balancing}. 
In more financial terms, the breakdown of the conditions for this 
dynamical equilibrium is, for
example, a {\it liquidity crisis}: a
sudden cooperativity of market orders, that lead to an increase of the trade 
sign correlation
function, can out-weight the liquidity providers stabilizing (mean-reverting) 
role, and lead to crashes. This
suggests that one should be able to write a mathematical model, 
inspired by our results, 
to describe this `on-off intermittency' scenario, advocated 
(although in a different context) in \cite{balancing,lux,holyst}. 
\vskip 0.5cm
\noindent{Acknowledgments:} {\small{Yuval Gefen thanks 
Science \& Finance/Capital Fund Management for hospitality during the
period this work was completed. We thank Jelle Boersma, Lisa Borland and Bernd Rosenow
for inspiring remarks, and J. Doyne Farmer for communicating his results 
on the autocorrelation of trades before publication \cite{Farmer3} and 
for insightful 
discussions and comments. We also thanks S. Picozzi 
for an interesting discussion and for pointing out the possible relevance 
of ref. \cite{balancing} to financial markets. Xavier Gabaix has made some 
crucial remarks on the first version of the manuscript, that in particular
lead to the material contained in the Appendix, and draw our attention to ref. \cite{Hopman}. 
Finally, we take the opportunity of
this paper to acknowledge the countless efforts of Gene Stanley to investigate the
dynamics of complex systems and to bring together different fields and ideas -- as testified
by the papers cited in reference, which inspired the present work.}}

\section*{Appendix: The case of a strictly diffusive process}

This appendix was inspired by a remark of Xavier Gabaix.
There is one particular case of our micro-model of prices, Eq. (\ref{model}), 
where prices are purely diffusive {\it at all times} 
(rather than only asymptotically). This is the case provided a
specific relation between the bare propagator $G_0$ and
the sign correlation function ${\cal C}_2(\ell)$ holds. 
In order to show this, let
us assume that the random variable 
$q_n \equiv \varepsilon_n \ln V_n$ can be written as:
\be\label{app1}
q_n =  \sum_{m \leq n} K(n-m) \xi_m,
\ee   
where $\xi_n$ are uncorrelated random variables ($\langle \xi_n \xi_m
\rangle = \langle \ln^2 V \rangle \delta_{n,m}$), and $K(.)$ a certain
kernel. In order for the $q_n$ to have the required correlations, the 
kernel $K(.)$ should obey the following equation:
\be
{\cal C}_2(n) = \langle \ln^2 V \rangle  \sum_{m \geq 0} K(m+n)K(m). 
\ee
In the case where ${\cal C}_2$ decays as $\ell^{-\gamma}$ with $0 < \gamma <1$,
it is easy to show that the asymptotic decay of $K(n)$ should also be 
a power-law $n^{-\delta}$ with $2 \delta -1 = \gamma$. Note that 
$1/2 < \delta < 1$. 

Inverting Eq. (\ref{app1}) allows one to obtain a set of uncorrelated 
random variables $\xi_n$ from a set of correlated variables $q_n$:
\be\label{app2}
\xi_n =  \sum_{m \leq n} Q(n-m) q_m,
\ee 
where $Q$ is the matrix inverse of $K$, 
such that $\sum_{m=0}^n K(n-m) Q(m)=\delta_{m,n}$. 
Eqs. (\ref{app1},\ref{app2}) 
in fact form the basis of linear filter theories, and $\xi_n$ can be 
seen as the prediction error on the next variable $q_n$. 
 
Introducing discrete Laplace transforms:
\be
\widehat{K}(E)=\sum_{n \geq 0} K(n) e^{-nE} \qquad 
\widehat{Q}(E)=\sum_{n \geq 0} Q(n) e^{-nE}, 
\ee
one finds $\widehat{K}(E)\widehat{Q}(E)=1$. For a power-law kernel $K(.)$,
one obtains: $\widehat{Q}(E) \propto E^{1-\delta}$ for $E \to 0$, 
and therefore $Q(n) \propto n^{\delta-2}$ for large $n$. It is useful 
to note that in this case $\widehat{Q}(E=0)=\sum_{n \geq 0} Q(n)=0$. 

Now, it is clear that if one defines the price process $p_n$ as:
\be
p_n = \sum_{m<n} \xi_m,
\ee
then $p_n$ is a diffusion process with a strictly
linear ${\cal D}(\ell)$, since the $\xi$'s are by construction uncorrelated. 
The price 
defined in this way can also be
written, using Eq. (\ref{app2}), as a linear combination of past $q_m$'s, as assumed in our
micro-model Eq. (\ref{model}), with:
\be\label{exact}
G_0^*(\ell) \equiv \sum_{m=0}^{\ell-1} Q(m).
\ee
This is an exact relation between ${\cal C}_2$ (that allows one to 
compute in turn $K$ and $Q$) and the response function $G_0^*$ for all 
$\ell$'s, where the
star indicates that strict diffusion is imposed.  

In the case
of power-law kernels, one finds from 
the above relation and from
$Q(n) \propto n^{\delta-2}$ for large $n$: 
\be
G_0^*(\ell) \propto \ell^{\delta - 1} \longrightarrow \beta = 1 -\delta = 
\frac{1-\gamma}{2},
\ee
which is, not surprisingly, 
the relation obtained in the main text from the assumption that 
prices are diffusive on long time scales. 

Eq. (\ref{exact}) can be used to construct $G_0^*$ from the empirical
determination of ${\cal C}_2$, shown in Fig. 10. In order to obtain 
this curve, we have fitted ${\cal C}_2(n)$ as:
\be
{\cal C}_2(0)=33.5; \qquad {\cal C}_2(n)=\frac{7.16}{(1.95+n)^{0.285}},
\ee
and used the Levinson-Durbin recursion algorithm for solving a Toeplitz system
(see, e.g., \cite{Percival}).


\end{document}